\documentclass[twocolumn,showpacs,preprintnumbers,amsmath,amssymb,prb,floatfix]{revtex4}

\usepackage{graphicx}
\usepackage{dcolumn}
\usepackage{bm}
\usepackage{color}

\begin{document}

\preprint{}

\title{Comment on `Spin Ice: Magnetic Excitations without Monopole Signature Using $\mu$SR'}

\author{S. T. Bramwell$^{1}$}
\email{s.t.bramwell@ucl.ac.uk}
\author{S. R. Giblin$^{2}$}
\email{sean.giblin@stfc.ac.uk}

\affiliation{
$^1$ London Centre for Nanotechnology, 
University College London, London WC1H 0AH, UK\\
$^2$ISIS Facility, Rutherford Appleton Laboratory, Chilton, Oxfordshire, OX11 0QX, U.K.} 
\date{\today}

\maketitle 


Dunsiger and co-workers (D {\it et al.})~\cite{DUGL} are wrong to contend that our experiment~\cite{Us} was `flawed in its conceptual design and execution and incorrect in its theoretical interpretation of the muon spin depolarization rate'. Their experiment does not compromise our conclusions~\cite{Us} and their deductions are contradicted by established facts. 
 
Muons implanted at typical sites within Dy$_2$Ti$_2$O$_7$ (DTO)~\cite{Lago, Us} experience local fields $\delta B \gg B_0$ (the applied field) which cause rapid dephasing of the muon precession. We analysed~\cite{Us} a second, minority, component with $\delta B\ll B_0$. The corresponding muon dephasing rate $\lambda(B_0,T)$ collapses to give the monopole charge $Q$.

Our interpretation relies on $\lambda\propto x$, the dimensionless monopole density and unique thermodynamic system variable. The correlation length of the monopole Coulomb gas is the Debye length, $l_D\propto x^{-1/2}$ and by dynamical scaling the magnetic relaxation rate $\nu \propto l_D^{-z}$, with $z=2$. Hence $\nu =\nu_0 x$ where $\nu_0$ can be shown to be the monopole hop rate.  Introducing a field scale $\Delta_0$, in the slow fluctuation limit, we find $\lambda = \Delta_0(\nu/\nu_0)^{y}$, with $y \approx 1$. Thus
$\lambda \propto \nu \propto x$ as we assumed. 
D~{\it et~al.}'s claim that our theory is flawed is based on their statement that $\lambda$ can never take this form, but they overlook the 
formal result of dimensional analysis that $\nu/\nu_0 = x(T,B_0)$ cannot be neglected in the limit $\nu/\Delta_0, \nu_0/\Delta_0 \rightarrow 0$. Within variations in $y$, to which our method is insensitive, our quoted formulae~\cite{Us} for $\lambda(\nu)$ are correct, but we apologise for misleadingly citing a theory (Ref. 21 in Ref. 2) that does not explicitly cover this case.  

D~{\it et~al.}'s second claim, based on analysis of a model, is that our signal arises from muons exterior to the sample. Our methodology~\cite{Giblin,Calder,ISIS} exploits both interior and exterior muons, and aims to separate near and far field contributions. Our recent experiments have shown that exterior muons in the silver backing plate dominate the signal of Ref. 2 at $T>0.4$ K where the Wien effect is absent, and measure the sample magnetization. The signal at $T <0.4$ K cannot be explained in this way, and correction of our data set for muons at distances greater than 0.25 mm from the sample brings it even closer to the theory of the Wien effect by removing the anomalous rise in $Q$ at $T > 0.3$ K (Fig. 5 of Ref. 2). 

Our evidence~\cite{Us} suggests that the Wien effect signal originates from interior muons, but sufficiently close exterior muons could achieve the same result  because $\lambda$ measures $x$ via the monopolar far field.  In fact, our experiment was first designed to measure the exterior response\cite{ISIS}. 


D~{\it et~al.} claim that our experiment is unreproducible, but they did not try to reproduce it. 
Instead they compared the effect of Ag and GaAs sample mounts at 0.1 K, 0.002 T (their Fig. 1). D~{\it et~al.}'s measurement had errors, both systematic (strong damping, poor thermal contact with GaAs) and random (poor statistics at $t > 6~\mu$s, large cryostat background) that would dominate 
the effect we study. In contrast, in early 2011, another group reproduced our result on a different spin ice system~\cite{Lees}. Similarly, we have confirmed that the Wien effect signal disappears when we rotate our sample so that muons are implanted directly into the silver backing plate, or when we use non-spin ice samples. 

If future $\mu$SR studies were to reveal a valid alternative interpretation of the minority response, we would, of course, revise our conclusions in the light of new information, but the results and criticisms of D~{\it et~al.} do not warrant any such revision.  In their Reply D~{\it et~al.} need to explain why they think our data collapse is reproducible, why it gives the correct $Q$ and why inferences of our result~\cite{Us} have been subsequently confirmed~\cite{Natphys}. 

D~{\it et~al.} finally observe that their experiment, like ours, is inconsistent with their model predictions. Rather than attributing this to known spin ice properties they invoke random crystal fields (unlikely for highly crystalline DTO) and unspecified `persistent spin dynamics'.  The latter are claimed to appear at temperatures ($\sim$ 4 K)  where single and double charge monopoles - or their equivalent classical spin flips - almost fully account for the specific heat~\cite{Hertog, CMS2} and ac-susceptibility~\cite{JH}. While non spin flip dynamics  are not excluded~\cite{Lago,Ehlers,E2}, the $\mu SR$ relaxation time in weak longitudinal field closely tracks that from ac-susceptibility as a function of temperature, but differs in the absolute value~\cite{Lago}. This implies a significant monopole contribution, contrary to D~{\it et~al.}'s claim. 	


\end{document}